\documentclass[aps,prb,showpacs,twocolumn,groupedaddress,floatfix]{revtex4-1}

\usepackage{amsmath,amssymb}
\usepackage{graphics}

\begin{document}

\title{Increasing thermoelectric performance using coherent transport}
\author{O. Karlstr\"om$^1$, H. Linke$^1$, G. Karlstr\"om$^2$, A. Wacker$^1$}

\affiliation{$^1$The Nanometer Structure Consortium (nmC@LU), Lund University, Box 118, SE-221 00, Sweden\\
$^2$Division of Theoretical Chemistry, Department of Chemistry, Lund University, Box 124, SE-221 00, Sweden}

\begin{abstract}
We show that coherent electron transport through zero-dimensional systems can be used to tailor the shape of the system's transmission function. 
This quantum-engineering approach can be used to enhance the performance of quantum dots or molecules in thermal-to-electric power conversion. 
Specifically, we show that electron interference in a two-level system can substantially improve the maximum thermoelectric power and the efficiency 
at maximum power by suppressing parasitic charge flow near the Fermi energy, and by reducing electronic heat conduction. We discuss possible 
realizations of this approach in molecular junctions or quantum dots.
\end{abstract}

\pacs{72.20.Pa,73.23.-b,85.35.Ds}

\maketitle

Thermoelectric devices are currently of high interest both for solid-state cooling and for increasing energy efficiency in converting
heat into electric power. The efficiency of thermoelectric materials
to convert a temperature gradient into electrical work is characterized by the
figure of merit \cite{GoldsmidBook1964}
\begin{eqnarray}
\label{ZT}
ZT=\frac{S^2\sigma T}{\kappa_{el}+\kappa_{ph}},
\end{eqnarray} 
with the temperature $T$, the Seebeck coefficient $S$, the electric conductance
$\sigma$,  as well as the electronic and phononic  thermal conductance
$\kappa_{el}$ and $\kappa_{ph}$. Other important performance parameters are the power output
$P=-(\mu_L-\mu_R)I$, the maximum power $P_{max}$ at the optimal bias and level configuration, and the
efficiency $\eta=P/J_Q$. Here, $\mu_L$ and
$\mu_R$ are the chemical potentials of left and right bath, respectively, $I$
is the particle current, and $J_Q$ is the heat flux out of the warm bath. 

Nanostructured materials are attractive candidates for efficient
thermoelectrics, because they offer the opportunity to optimize $S$
by using the energy-selectivity of charge carrier transport in
low-dimensional systems \cite{DresselhausAdvMat2007} combined with the
suppression of  $\kappa_{ph}$ by interface scattering.\cite{WangScience2007} In such systems, $I,S,\sigma,J_Q$ and $\kappa_{el}$ can be directly
evaluated from the electronic transmission function $\Sigma(E)$.\cite{Mahan1996}

In particular, zero-dimensional (0D) systems, such as quantum dots or
molecules, that are weakly coupled to electron reservoirs,  can be designed as
ideal energy filters with an energy-dependent transmission function
$\Sigma(E)\propto \delta(E-E_1)$,  where $E_1$ is the position of the single
level that contributes to transport. In this limit, and for  $\kappa_{ph}  = 0
$,  $ZT$ diverges \cite{Mahan1996} and the efficiency of thermoelectric power
conversion approaches Carnot efficiency ($\eta _C$).\cite{HumphreyPRL2002,VandenBroeckACP2007} Additionally, such ideal quantum dots reach the
ideal Curzon-Ahlborn efficiency $\eta_{CA}$ of about  $\eta _C/2$ when $E_1$
is tuned to maximum  power production $P_{max}$.\cite{EspositoEPL2009} However, the limit $\Sigma(E)\propto\delta(E-E_1)$ in 0D systems
is not interesting for applications because the power output
becomes exceedingly small,
and even a small $\kappa_{ph}\neq 0$ leads to a low value of $ZT$.
To increase the current and $P_{max}$ in 0D systems one needs to broaden
$\Sigma(E)$. This drastically reduces the efficiency at maximum
power $\eta_{max P}$, because the Lorentzian-shaped
$\Sigma(E)$ has a long low-energy tail that leads to a counter flow of cold
charge carriers, contributing with opposite sign to $S$.\cite{NakpathomkunPRB2010}

Here, we show that coherent transport in 0D systems can be used to tailor $\Sigma(E)$ 
such that counter flow of cold charge carriers is effectively suppressed. In
addition to its fundamental interest, 
this approach to quantum engineering is shown to substantially increase $P_{max}$,
$\eta_{max P}$ and $ZT$ compared to the ideal 0D systems addressed above. 
We discuss how this effect can be implemented in semiconductor quantum dots or in
molecular junctions.  

\begin{figure}[t]
\begin{center}
\begin{minipage}{.25\linewidth}
\resizebox{!}{15mm}{\includegraphics{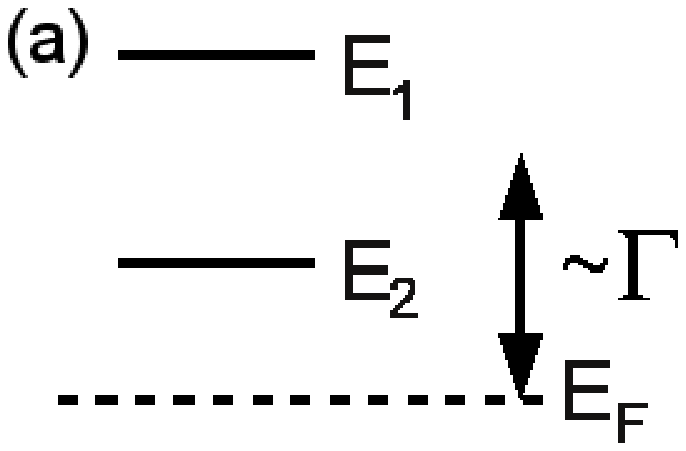}}
\end{minipage}
\begin{minipage}{.33\linewidth}
\resizebox{!}{20mm}{\includegraphics{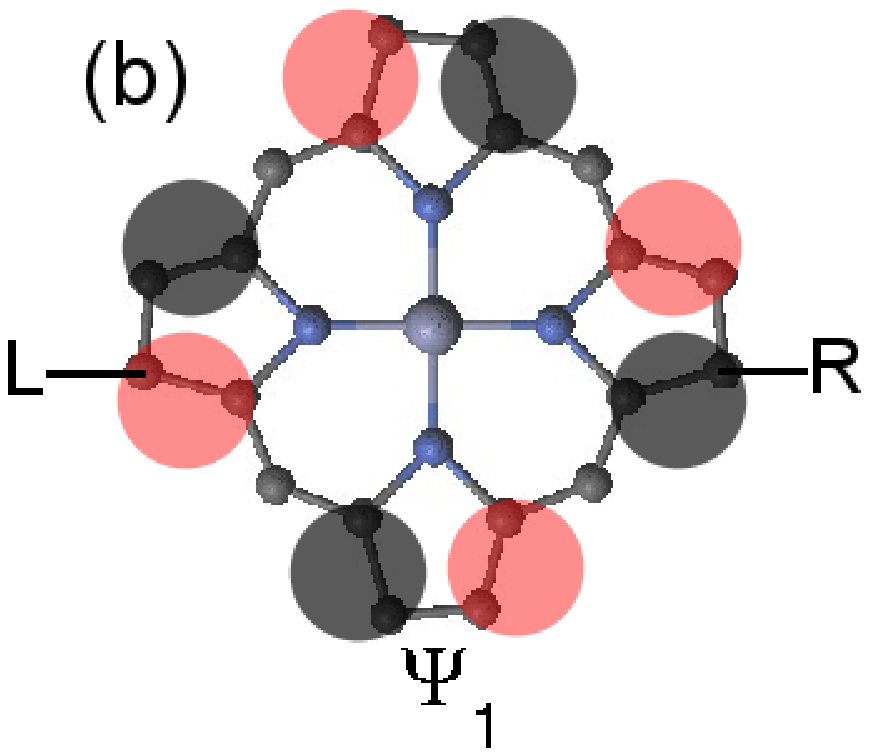}}
\end{minipage}
\begin{minipage}{.33\linewidth}
\resizebox{!}{20mm}{\includegraphics{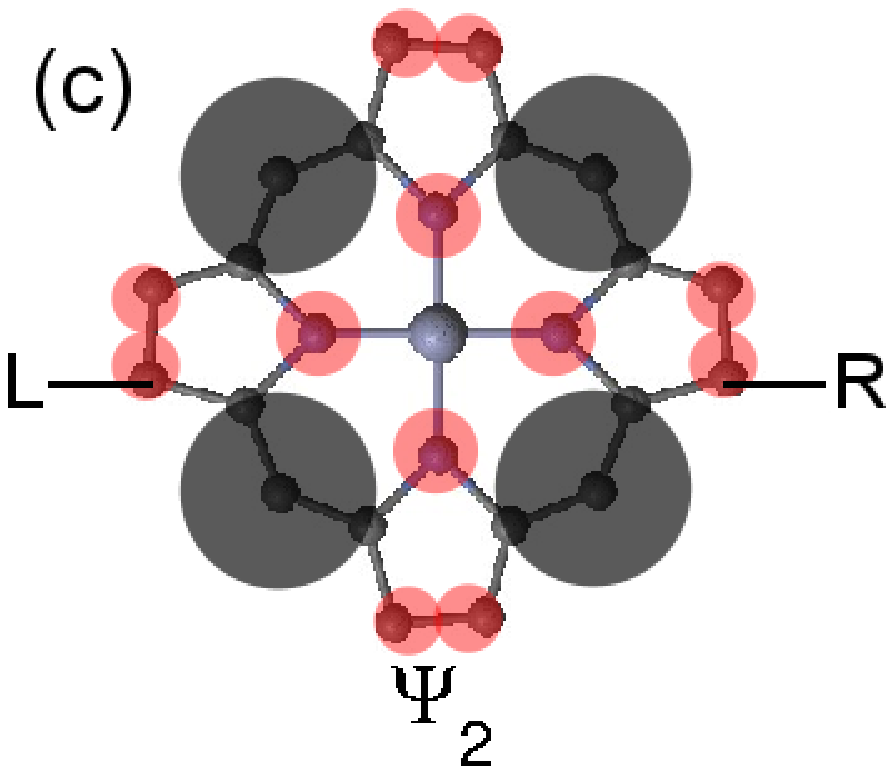}}
\end{minipage}
\end{center}
\caption{(a)~Level positions in the proposed two-level system. (b)-(c)~Possible implementation using zinc porphine,
sketching the wave functions of the two almost degenerate levels close to the Fermi energy, see also Ref.~\onlinecite{Tsaicpl2002}.
The light/dark shaded regions indicate a
positive/negative amplitude of the wave function $\Psi$, respectively.
By replacing the hydrogen atoms at the positions L and R with, for example, 
sulfur the molecule can be contacted to leads so that the parity of the wave functions with respect to the leads differ for the two levels.}
\label{Fig1}
\end{figure}

We consider a two-level system with both energy levels $E_1, E_2$ situated on one
side of $\mu_L$ and $\mu_R$, (Fig.~\ref{Fig1}(a)),
such that the charging energy is of minor 
importance,\cite{KarlstromPRB2011} and a spin degeneracy will not affect transport significantly.
This setup is similar to the double-dot 
case\footnote{P. Trocha and J. Barna{\'s}, arXiv:1108.2422v1; G. G{\'o}mez-Silva, O.{\'A}valos-Ovando, M. L. Ladr{\'o}n de Guevara and P. A. Orellana, arXiv:1108.4460v1.} 
addressed very recently in Ref.~\onlinecite{LiuNT2011}, where equal coupling strengths of the two levels were assumed.
Related findings were presented in Ref.~\onlinecite{WierzbickiPRB2011} where the thermoelectric properties of double quantum dots with couplings ranging from
serial to parallel configurations were investigated, assuming equal energies of the two dots.

The coupling between lead and dot is parametrized as in Ref.~\onlinecite{KarlstromPRB2011}, assuming equal coupling strengths to left and
right leads. The two levels couple with different parity to the leads and their coupling strengths differ by a factor $a^2$,
$\Gamma_{L1}=\Gamma_{R1}=\Gamma,~\Gamma_{L2}=\Gamma_{R2}=a^{2}\Gamma$. This difference in parity will turn out to be essential for the
increased thermoelectric performance.
For such couplings it can be shown that the Breit-Wigner formula provides results identical to 
the exact method of nonequilibrium Green's functions (NEGF) \footnote{This definition of $\Gamma$ differs by a factor $2$ from the one used in Ref.~ 
\onlinecite{NakpathomkunPRB2010}.}

\begin{eqnarray}
\label{BW}
\Sigma(E)=\Gamma^{2}\left|\frac{1}{E-E_1+i\Gamma}-\frac{a^{2}}{E-E_2+ia^{2}\Gamma}\right|^{2},
\end{eqnarray}

\noindent where the two levels contribute with different sign due to the opposite parity. Vanishing transport at the Fermi energy $E_F\equiv0$ 
corresponds to $E_2=a^{2}E_1$. A device operated close to such a level configuration would have the desired property that 
$\Sigma(0)\approx0$ independently of $\Gamma$. 

\begin{figure}[ht]
\begin{center}
{\resizebox{!}{39mm}{\includegraphics{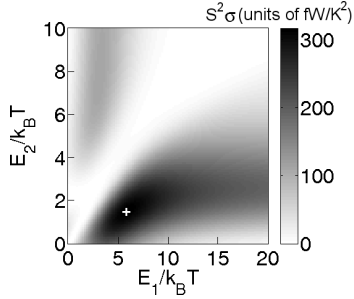}}}
\end{center}
\caption{The power factor for coherent transport through a quantum system with two levels and for $T=300$~K, $\Gamma=3k_BT_R$, and $a=0.6$. 
Level positions are defined relative to $E_F$. 
The cross marks the maximum ($S^2\sigma=316$~fW at $E_1=5.80k_BT$, $E_2=1.47k_BT$), 
and the transmission function is
calculated at this level configuration in Fig.~\ref{Pmax}(a).}
\label{power2}
\end{figure}

\begin{figure}[ht]
\begin{center}
\begin{minipage}{.45\linewidth}
\resizebox{!}{37mm}{\includegraphics{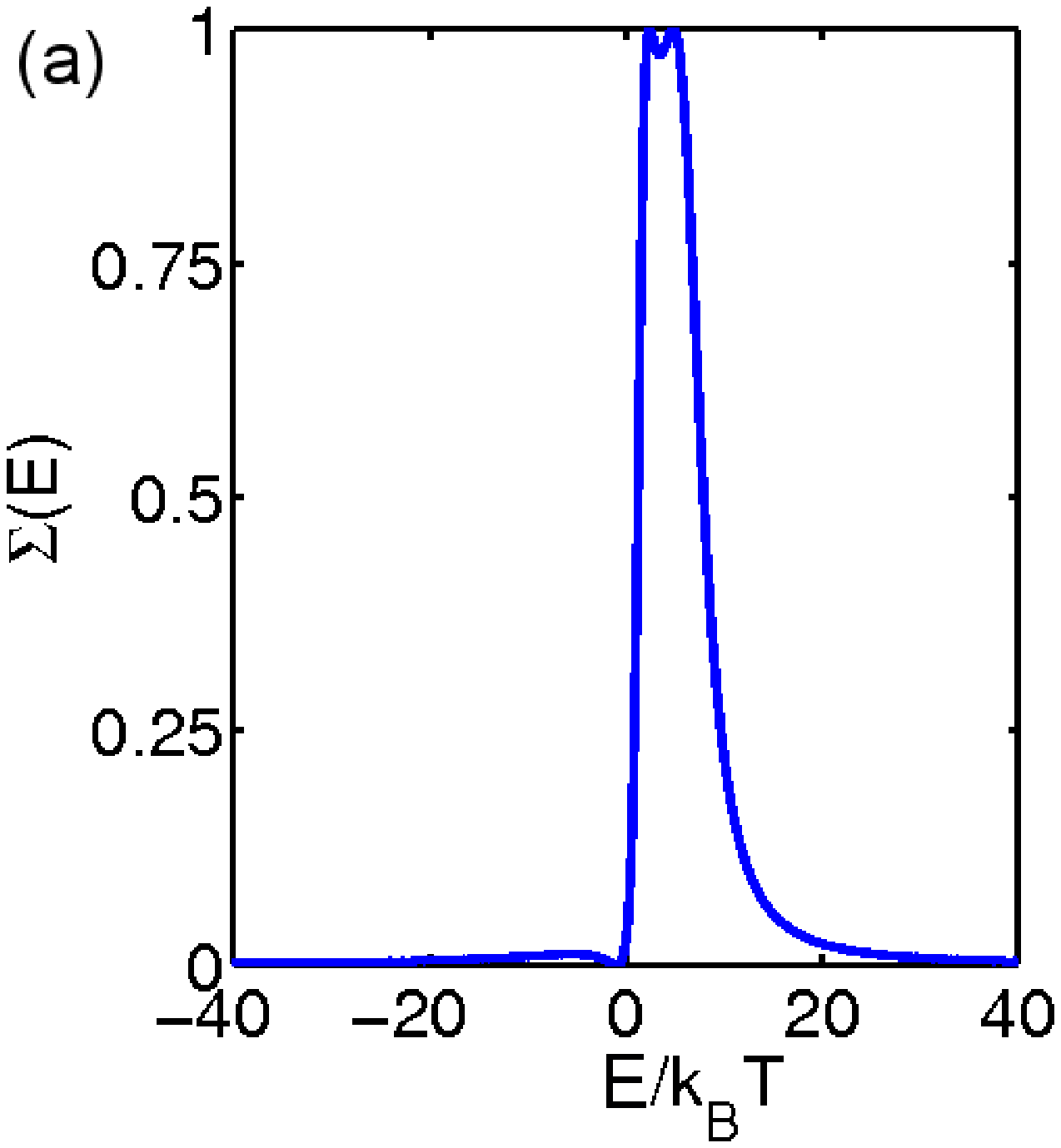}}
\end{minipage}
\begin{minipage}{.45\linewidth}
\resizebox{!}{37mm}{\includegraphics{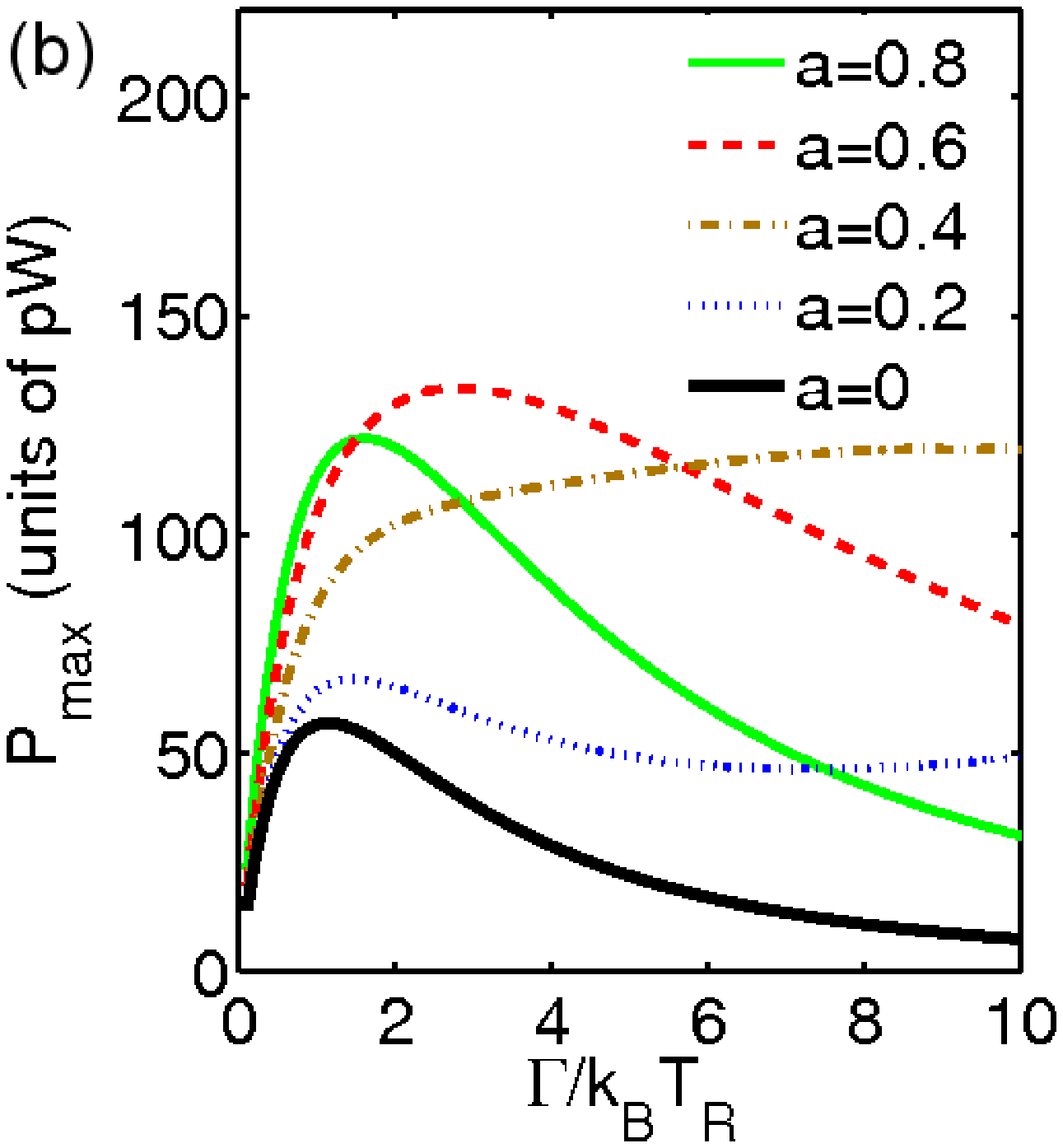}}
\end{minipage}
\end{center}
\begin{center}
\begin{minipage}{.45\linewidth}
\resizebox{!}{37mm}{\includegraphics{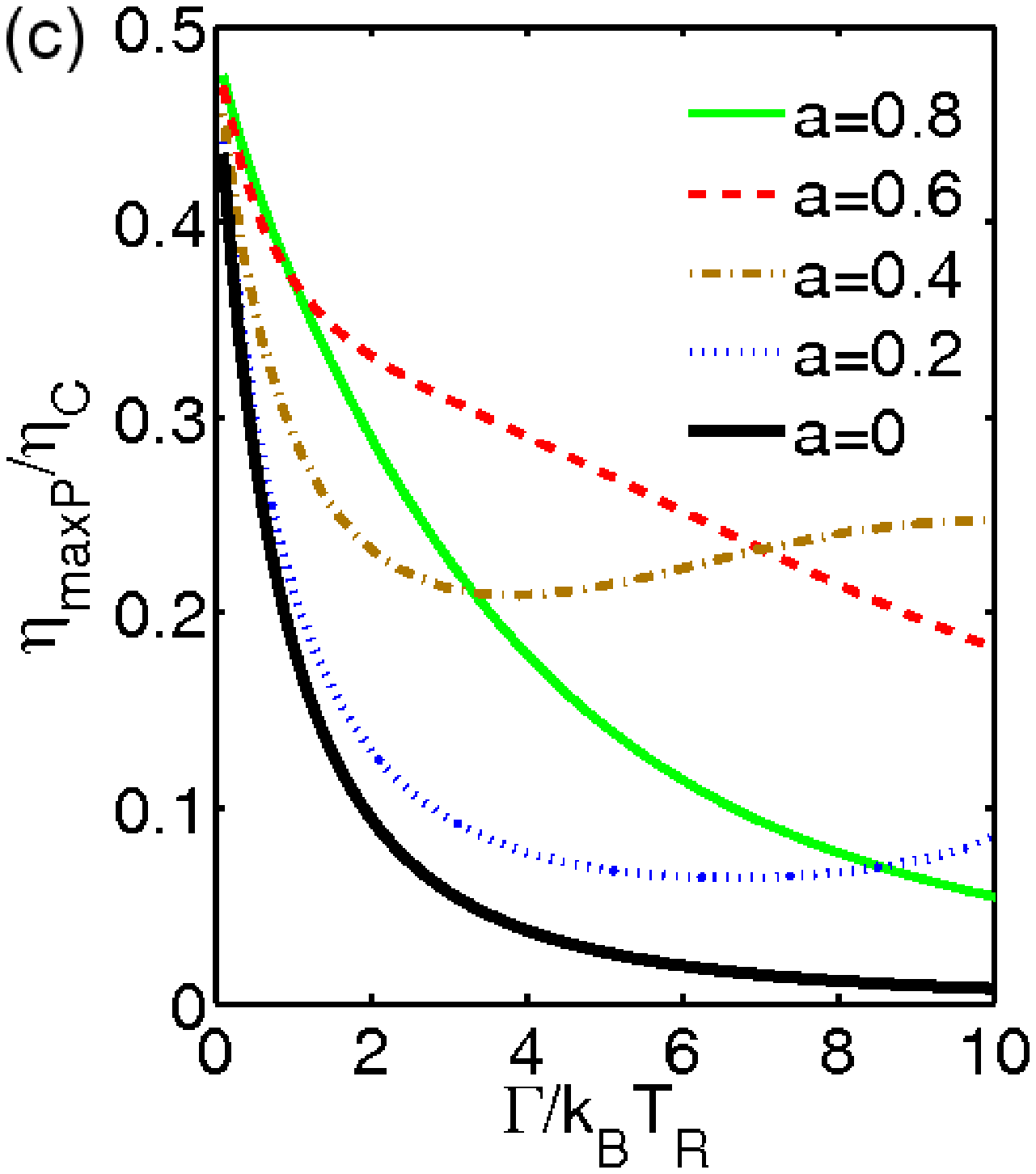}}
\end{minipage}
\begin{minipage}{.45\linewidth}
\resizebox{!}{37mm}{\includegraphics{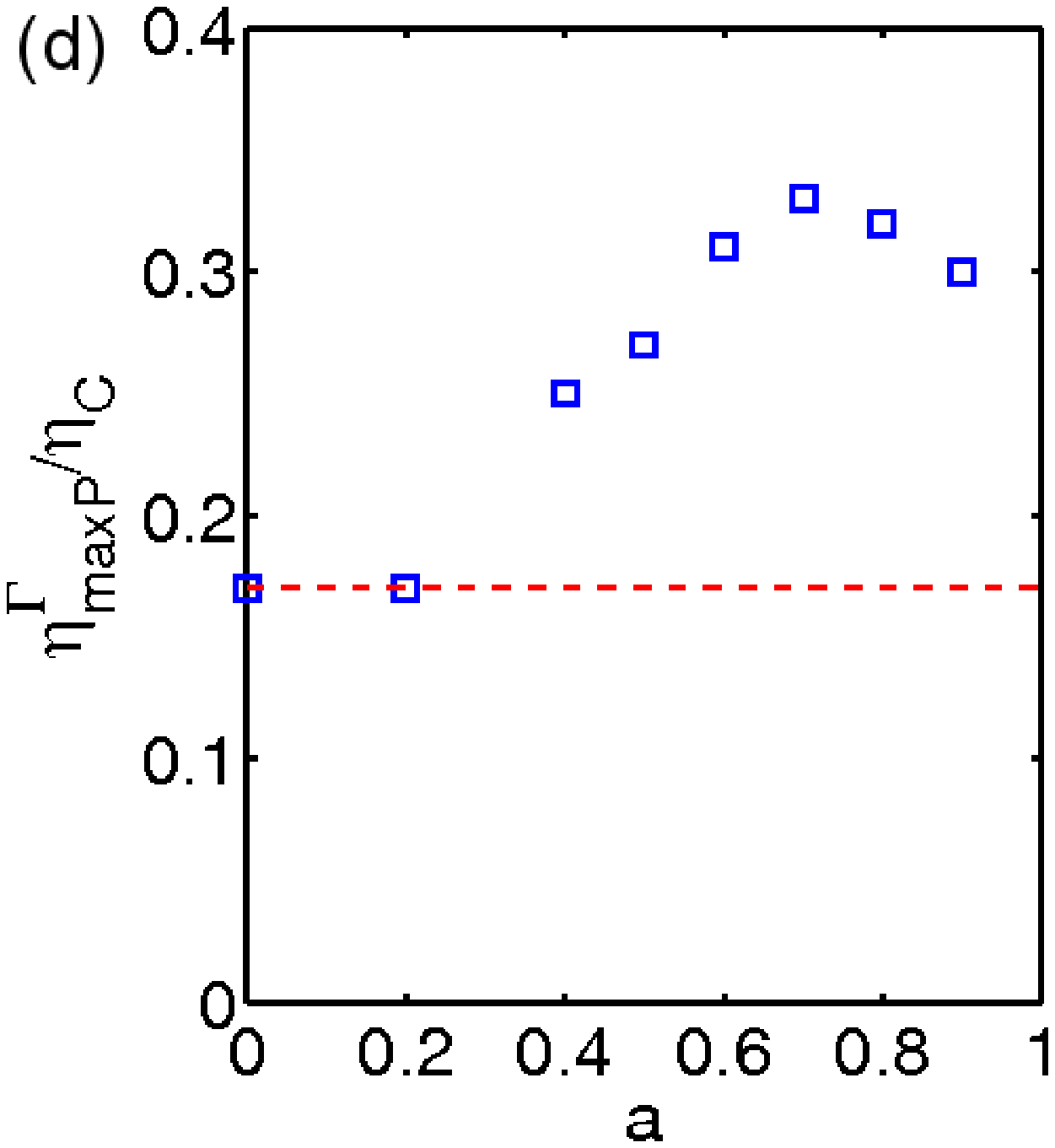}}
\end{minipage}
\end{center}
\caption{(a)~Transmission function at the level configuration corresponding to maximum power, see Fig.~\ref{power2}.
(b)~Maximum power production and (c)~efficiency at maximum power $\eta_{maxP}$ in units of the Carnot efficiency $\eta_C$, as a function of $\Gamma$. 
(d)~shows $\eta_{maxP}^{\Gamma}$ (defined as $P_{max}$ at the optimal choice of $\Gamma$, see text), with the dashed line marking the one-level result.
In (b)-(d) the temperatures of the two leads are given by $T_L=330$ K and $T_R=300$ K.}
\label{Pmax}
\end{figure}

Fig.~\ref{power2} shows that a large power factor $S^2\sigma$ is achieved for level positions over an energy range 
as large as several $k_BT$, ($k_BT\approx0.26$~eV at $T=300$~K).
The transmission function at maximum power factor, which is marked by a cross in Fig.~\ref{power2},
is displayed in Fig.~\ref{Pmax}(a). 
The strong asymmetry in $\Sigma(E)$, with a sharp step facing $E_F$, is ideal
for high power production.\cite{Humphrey2005} The decrease of $\Sigma(E)$ for large energies prevents the transmission of high-energy electrons, 
which reduces $\kappa_{el}$ and results in an increased efficiency of the device.

Fig.~\ref{Pmax}(b) displays $P_{\mathrm{max}}(\Gamma)$ for different values of the 
asymmetry parameter $a$, showing peak power values more than two times larger than the $P_{max}$ achievable in a single-level 0D system corresponding to $a=0$.
In the coherent case maximum power occurs for larger $\Gamma$, 
since the broadening of the levels is not such a severe problem. Fig.~\ref{Pmax}(c) shows the efficiency at maximum power $\eta_{maxP}$, 
which can be increased by over an order of magnitude, compared to the single level case, for large $\Gamma$.
In the limit $\Gamma\rightarrow0$ one observes that $\eta_{maxP}\rightarrow\eta_{CA}\approx0.5\eta_C$.\cite{EspositoEPL2009,NakpathomkunPRB2010} 
When $\Gamma$ and $a$ are small $P_{max}$ and
$\eta_{maxP}$ are not substantially increased, since the effect of the second level is small when $k_BT\gg a^2\Gamma$. As $\Gamma$ is increased, the effect
of the second level can be observed as an increase in $P_{max}$ and $\eta_{maxP}$, which can result in a local maximum.
Fig.~\ref{Pmax}(d) displays $\eta_{maxP}^{\Gamma}$, 
the efficiency at maximum power where the power production is optimized with respect to $\Gamma$ as well as with respect to bias and level positions. 
We restrict ourselves to
$\Gamma<10k_BT_R$ where effects of the local maximum, present for small $a$, is not observed.
For small $a$ the presence of the second level is negligible, and the efficiency of a single level is approached.
High efficiency is achieved close to $E_2=a^2E_1$. For $a\rightarrow1$ the level configuration that yields $P_{max}$ does not coincide with $E_2=a^2E_1$ where 
transport is blocked (Eq.~\ref{BW}).
Thus $\eta_{maxP}^{\Gamma}$ decreases for large $a$ and the maximum is found around $a=0.7$.

\begin{figure}[ht]
\begin{center}
\begin{minipage}{.49\linewidth}
\resizebox{!}{36mm}{\includegraphics{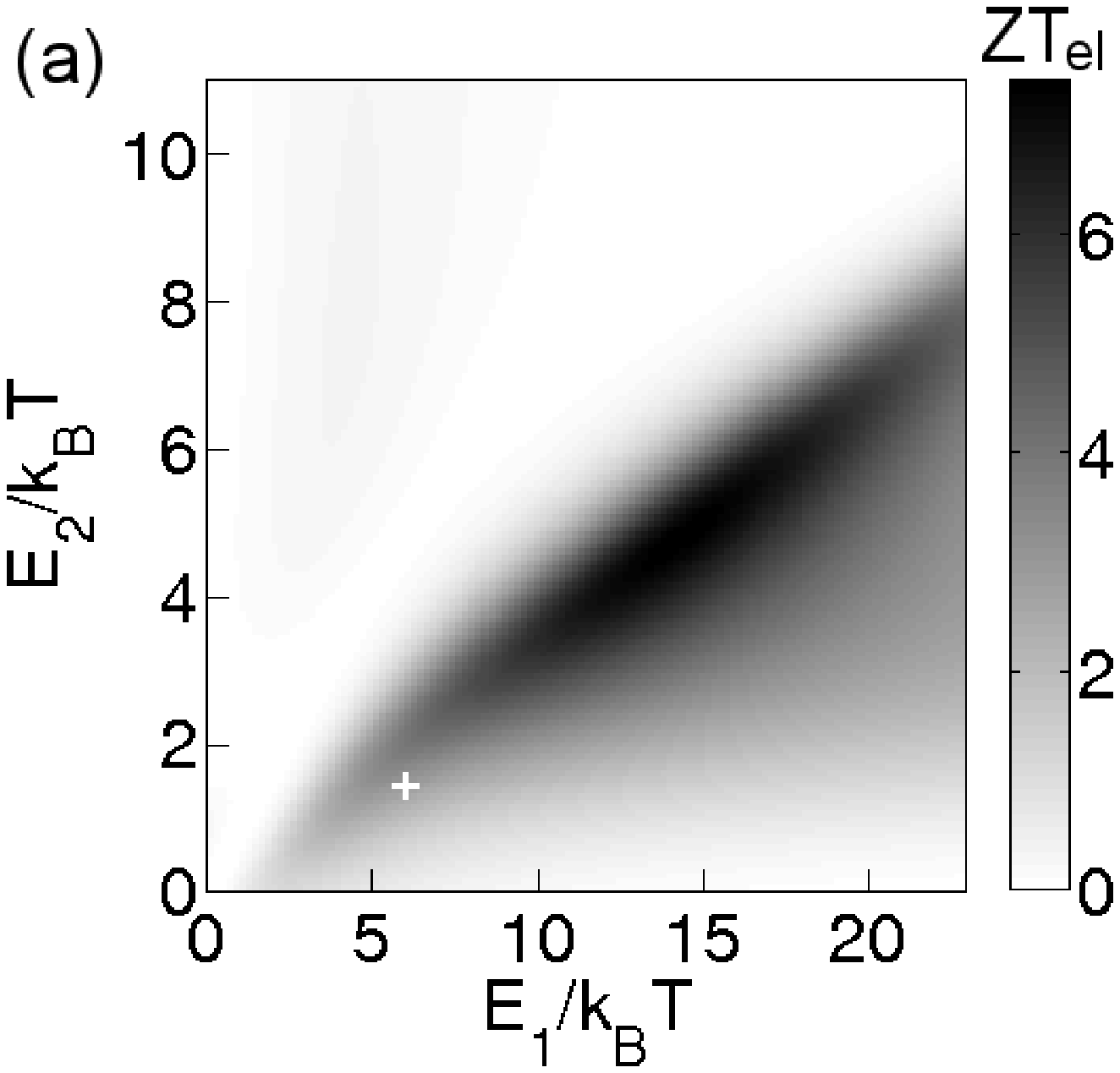}}
\end{minipage}
\begin{minipage}{.49\linewidth}
\resizebox{!}{36mm}{\includegraphics{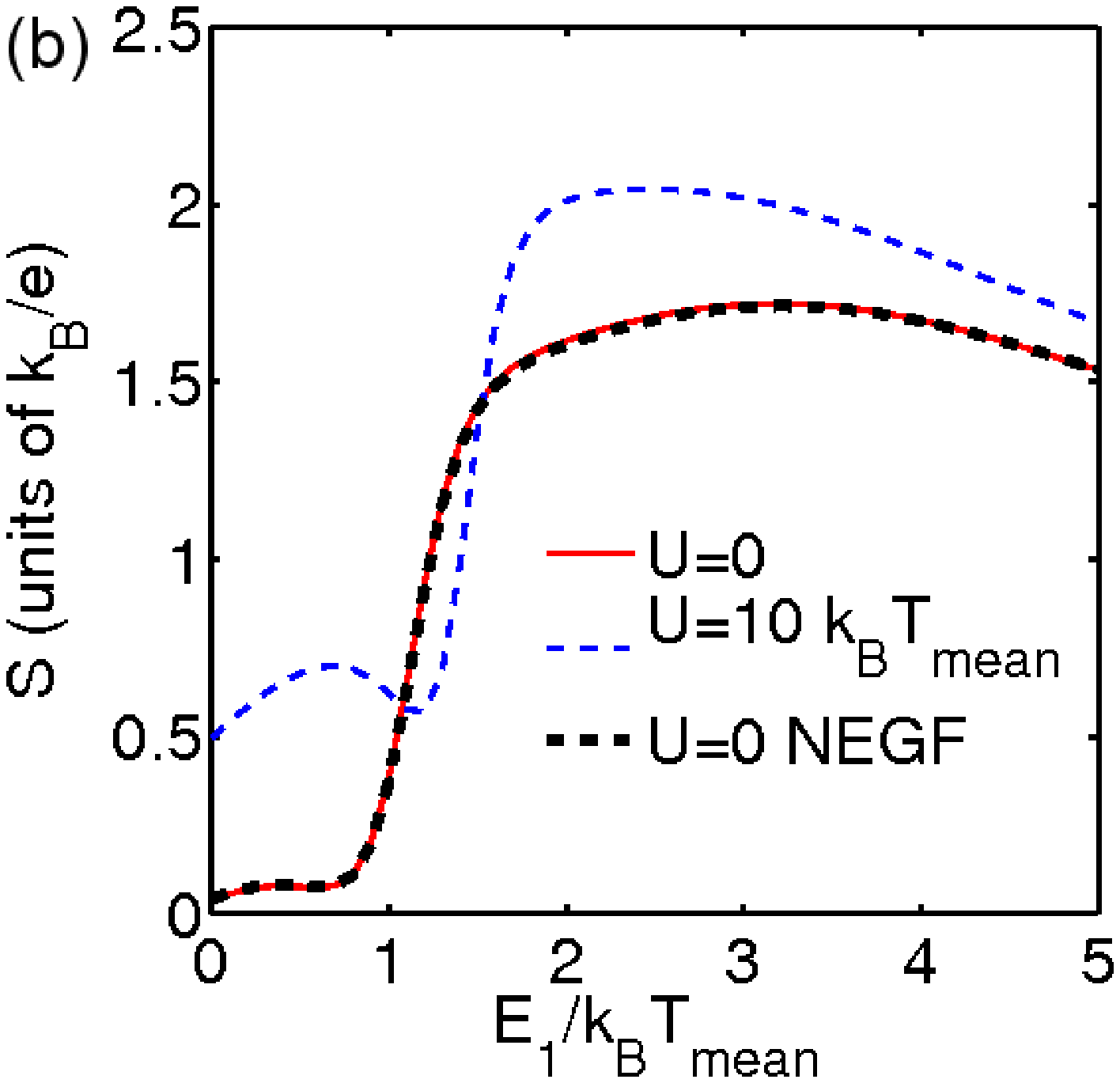}}
\end{minipage}
\end{center}
\caption{(a)~Thermoelectric figure of merit $ZT_{el}$ for coherent transport as a function of level configuration
for $T=300$~K. The parameters 
$\Gamma=3k_BT$, and $a=0.6$ match the conditions for maximum power in Fig.~\ref{Pmax}(b). (b)~The thermopower $S$ calculated 
for $U=0$ and $U=10~k_BT_{\mathrm{mean}}$, $\Gamma=k_BT_{\mathrm{mean}}$, $a=0.7$, and $\Delta T=0.1~k_BT_{\mathrm{mean}}$. The second level is positioned at 
$E_2=k_BT_{\mathrm{mean}}$.}
\label{ZTfig}
\end{figure}

To facilitate the comparison with other work we present, in Fig.~\ref{ZTfig}(a), the resulting figure of merit $ZT_{el}$, defined by $\kappa_{ph}=0$ in 
Eq.~(\ref{ZT}).
We see that even for the relatively large $\Gamma$ resulting in maximum power, values as high as $ZT_{el}=7$ can be reached. 
In part, this increase is due to a decrease in $\kappa_{el}$. 
It is worth noting that, for the realistic case of finite $\kappa_{ph}$, maximum $ZT$ is expected near the conditions for
$P_{max}$ (indicated by crosses in Figs.~\ref{power2} and \ref{ZTfig}(a)), 
because a high power factor is needed in Eq.~(\ref{ZT}) to provide robustness against parasitic heat flow due to phonons.

Coherent transport through molecules has been discussed previously in 
Refs.~\onlinecite{CardamoneNanoLett2006,KeNanoLett2008}, where
the interference between levels positioned on different sides of $E_F$ causes a dip in the transmission function at $E_F$. The corresponding
reduction of charge flow near $E_F$ can be used to increase the 
thermoelectric figure of merit, for example in molecules \cite{BergfieldNanoLetters2009,*BergfieldACSNano2010} 
or quantum dots.\cite{Nakanishi_arXiv2007} 
However, the increased values of $ZT_{el}$ obtained in these papers are a result of the decrease in $\kappa_{el}$.
This results in low power output and in a high sensitivity to phonons \cite{BergfieldACSNano2010} because $ZT=ZT_{el}/(1+\kappa_{ph}/\kappa_{el})$.
Increased thermopower due to interference effects has also been discussed in the context of side-group induced Fano resonances.\cite{FinchPRB2009}
While the thermoelectric performance was significantly improved, this approach is challenging because it requires very good control 
($\sim10\,^{\circ}$) of the tilting angle of the side group. Recently, efforts were made to determine which types of molecules exhibit 
Fano resonances,\cite{MarkussenNL2010,*MarkussenPCCP2011} and investigate their thermoelectric performance.\cite{StadlerarXiv}

We now discuss possible experimental implementations of coherent enhancement of thermoelectric efficiency:\\
First, highly controllable 0D systems can be realized in the form of quantum dots defined in two-dimensional electron gases \cite{ScheibnerPRL2005} 
or in nanowires.\cite{HoffmannNanoLett2009,NilssonPRL2010}
By electrostatically gating the quantum dot, its energy levels can be shifted so that mainly two states contribute to the transport.
The relative position of the two levels with different parity can be controlled by applying a magnetic field,\cite{NilssonPRL2010}
in particular in quantum dot materials with a large g-factor such as InSb.\cite{NilssonNL2009}
The use of two parallel coupled quantum dots allows for better control of
the two levels, but coherence is decreased due to the spatial separation of the two dots.

The exquisite experimental control over semiconductor quantum dots thus makes them an excellent testbed for the fundamental study of the effects predicted here. 
However, the use of quantum dots in applications, for example in the form of nanocrystals embedded into bulk materials, 
is limited because it is very challenging to produce such dots with the high uniformity required to have equal energy levels in all dots.

An alternative system with inherently much better reproducibility are molecular junctions that fulfill the following requirements:\\
i) The molecule must have two almost degenerate, conductive states with different parity. 
Such states can be found in molecules that are symmetric with respect to the leads, where
the symmetry implies that the states must be symmetric and anti-symmetric.\\
ii) The two states must be positioned within a few $\Gamma$ of $E_F$ (Fig.~\ref{power2}).
The active states could be either LUMO-states which should be
placed slightly above $E_F$, or HOMO-states which should be placed slightly below.  

A possible candidate is zinc porphine (Fig.~\ref{Fig1}(b,c)). This
molecule has two almost degenerate HOMO-levels with the desired parity.\cite{Tsaicpl2002} The relative position of the two levels can be 
changed by adding e.g. phenyl groups \cite{WalshTheochem2006} or replacing all hydrogen atoms with e.g. chlorine or boron.\cite{Tsaicpl2002} 
To contact the molecule by gold electrodes one could replace two hydrogen atoms by sulfur (Fig.~\ref{Fig1}(b,c)). 
The leads can affect the molecular orbitals via charge 
transfer \cite{XueJCP2001} and symmetry breaking. To include such effects, and calculate the transmission function of the
system, one can use the method described in Ref.~\onlinecite{XueCP2002}. A rough estimate of the coupling between lead and orbital $i$ 
of the junction molecule is given by $\Gamma_i\approx2\pi|\Psi_i|^2C^2N_0$,
where $\Psi_i$ is the value of the wave function at the sulfur, $C=2$~eV is the matrix element between the gold and the sulfur atoms, 
and $N_0=0.07/$eV is the gold density of states per atom per eV at $E_F$.\cite{TianJCP1998} 
Assuming that the wave function is evenly distributed among the 20 carbon, 4 nitrogen, 1 zinc, and 2 sulfur atoms of zinc porphine gives
$\Psi_i=(27)^{-1/2}$ and $\Gamma_i\approx0.065$~eV~$\approx 2.5~k_BT$ at $T=300$~K. The actual values for $\Gamma_i$ will vary between the different
orbitals, but this estimate shows that the parameters are in a regime where coherent transport can be significantly more efficient than 
transport through a single level, (Fig.~\ref{Pmax}(b) and (c)).

We now turn to the effect of phonons on thermoelectric performance.
Values around $\kappa_{ph}=50$~pW/K have been experimentally determined for molecular wires,\cite{WangScience2007} 
and $\kappa_{ph}=10...100$~pW/K has been estimated for molecular junctions.\cite{MingoPRB2006} 
For the parameters of Fig.~\ref{ZTfig}(a) the latter results in impressive
$ZT=0.7...3$. These values are quite preliminary as the exact value of $\kappa_{ph}$ depends on the junction molecule. However, it should be noted that 
experiments and theoretical calculations of the IR spectra of
zinc porphine show that the majority of the vibrational transitions lie significantly above $k_BT$ at room temperature,\cite{Jarzecki1997} suggesting 
a low $\kappa_{ph}$.
Also any destructive effect of phonons on the coherence should not be a severe problem as the suggested operation point of the device 
is in the regime $\Gamma>k_BT$, i.e. the coherence time of the electron on the junction molecule is longer
than the tunneling time, resulting in coherent transport.
A more detailed study of the couplings between the low-energy vibrations and the electronic degrees of freedom is required. 

So far we have not discussed the effects of electron-electron interaction.
In the following we show that a finite charging energy $U$ does
not affect the results qualitatively, and may in fact result in an increased $S$. To include a finite $U$ we use the second order
von Neumann approach (2vN),\cite{PedersenPRB2005a} an equation of motion technique where co-tunneling as well as the coherence and 
charging energy between the levels are included. 
This approach calculates the current without directly calculating the transmission 
function $\Sigma(E)$, and $S$ is calculated as the open-circuit voltage at a small but finite temperature gradient $\Delta T$
applied between the leads.

The 2vN-approach neglects certain tunneling events of
higher order which results in less accurate results for lower temperatures. 
In Fig.~\ref{ZTfig}(b) the thermopower is calculated
for the temperatures 
$k_BT_L=k_BT_{\mathrm{mean}}+\Delta T/2$ and $k_BT_R=k_BT_{\mathrm{mean}}-\Delta T/2$. To investigate the importance of a 
finite $U$, $E_2$ has been placed
slightly above $E_F$ at $E_2=k_BT_{\mathrm{mean}}$. The thermopower is then calculated as a function of $E_1$. 
To demonstrate that the 2vN 
method gives accurate results, and that the effect of the finite $\Delta T$ is negligible, the $U=0$ results are compared to results
obtained using NEGF.
When $E_1\approx 0$ the transport
is dominated by this level, which results in a low thermopower in the case of $U=0$. For a finite $U$ part of the spectral density of
$E_1$ is shifted to $E_1+U$ due to the small but finite occupation of level $E_2$. This results in transport at increased 
energies, and consequently a higher thermopower. When $E_1$ is far away from $E_F$ the difference between $U=0$ and 
$U=10~k_BT_{\mathrm{mean}}$ decreases as the effect of the charging energy vanishes when the levels are empty.

In conclusion, we have shown how coherent transport can be used to tailor the shape of the system's transmission function, to combine a large 
thermoelectric power production
with a high efficiency. Compared to transport through one-level systems, the maximum power production is more than doubled. At the same time
efficiencies at maximum power up to $\eta_{maxP}=0.33$ are achieved for $\kappa_{ph}=0$, comparable to those theoretically achievable in ideal (ballistic) 
one-dimensional devices where $\eta_{maxP}=0.36$ is expected.\cite{NakpathomkunPRB2010} 
We also proposed a specific molecular junction for the implementation of this novel approach to thermoelectric engineering, 
and expect a ZT of the order of unity or above. 

We thank Christian Bergenfeldt and Peter Samuelsson for stimulating discussions. Financial support from the Swedish Research Council (VR), Energimyndigheten
(Grant No. 32920-1), and nmC@LU is gratefully acknowledged.


\begin{thebibliography}{35}%
\makeatletter
\providecommand \@ifxundefined [1]{%
 \@ifx{#1\undefined}
}%
\providecommand \@ifnum [1]{%
 \ifnum #1\expandafter \@firstoftwo
 \else \expandafter \@secondoftwo
 \fi
}%
\providecommand \@ifx [1]{%
 \ifx #1\expandafter \@firstoftwo
 \else \expandafter \@secondoftwo
 \fi
}%
\providecommand \natexlab [1]{#1}%
\providecommand \enquote  [1]{``#1''}%
\providecommand \bibnamefont  [1]{#1}%
\providecommand \bibfnamefont [1]{#1}%
\providecommand \citenamefont [1]{#1}%
\providecommand \href@noop [0]{\@secondoftwo}%
\providecommand \href [0]{\begingroup \@sanitize@url \@href}%
\providecommand \@href[1]{\@@startlink{#1}\@@href}%
\providecommand \@@href[1]{\endgroup#1\@@endlink}%
\providecommand \@sanitize@url [0]{\catcode `\\12\catcode `\$12\catcode
  `\&12\catcode `\#12\catcode `\^12\catcode `\_12\catcode `\%12\relax}%
\providecommand \@@startlink[1]{}%
\providecommand \@@endlink[0]{}%
\providecommand \url  [0]{\begingroup\@sanitize@url \@url }%
\providecommand \@url [1]{\endgroup\@href {#1}{\urlprefix }}%
\providecommand \urlprefix  [0]{URL }%
\providecommand \Eprint [0]{\href }%
\providecommand \doibase [0]{http://dx.doi.org/}%
\providecommand \selectlanguage [0]{\@gobble}%
\providecommand \bibinfo  [0]{\@secondoftwo}%
\providecommand \bibfield  [0]{\@secondoftwo}%
\providecommand \translation [1]{[#1]}%
\providecommand \BibitemOpen [0]{}%
\providecommand \bibitemStop [0]{}%
\providecommand \bibitemNoStop [0]{.\EOS\space}%
\providecommand \EOS [0]{\spacefactor3000\relax}%
\providecommand \BibitemShut  [1]{\csname bibitem#1\endcsname}%
\let\auto@bib@innerbib\@empty
\bibitem [{\citenamefont {Goldsmid}(1964)}]{GoldsmidBook1964}%
  \BibitemOpen
  \bibfield  {author} {\bibinfo {author} {\bibfnamefont {H.~J.}\ \bibnamefont
  {Goldsmid}},\ }\href@noop {} {\emph {\bibinfo {title} {Thermoelectric
  Refrigeration}}}\ (\bibinfo  {publisher} {Plenum Press},\ \bibinfo {address}
  {New York},\ \bibinfo {year} {1964})\BibitemShut {NoStop}%
\bibitem [{\citenamefont {Dresselhaus}\ \emph {et~al.}(2007)\citenamefont
  {Dresselhaus}, \citenamefont {Chen}, \citenamefont {Tang}, \citenamefont
  {Yang}, \citenamefont {Lee}, \citenamefont {Wang}, \citenamefont {Ren},
  \citenamefont {Fleurial},\ and\ \citenamefont
  {Gogna}}]{DresselhausAdvMat2007}%
  \BibitemOpen
  \bibfield  {author} {\bibinfo {author} {\bibfnamefont {M.~S.}\ \bibnamefont
  {Dresselhaus}}, \bibinfo {author} {\bibfnamefont {G.}~\bibnamefont {Chen}},
  \bibinfo {author} {\bibfnamefont {M.~Y.}\ \bibnamefont {Tang}}, \bibinfo
  {author} {\bibfnamefont {R.~G.}\ \bibnamefont {Yang}}, \bibinfo {author}
  {\bibfnamefont {H.}~\bibnamefont {Lee}}, \bibinfo {author} {\bibfnamefont
  {D.~Z.}\ \bibnamefont {Wang}}, \bibinfo {author} {\bibfnamefont {Z.~F.}\
  \bibnamefont {Ren}}, \bibinfo {author} {\bibfnamefont {J.-P.}\ \bibnamefont
  {Fleurial}}, \ and\ \bibinfo {author} {\bibfnamefont {P.}~\bibnamefont
  {Gogna}},\ }\href@noop {} {\bibfield  {journal} {\bibinfo  {journal}
  {Adv.~Mater}\ }\textbf {\bibinfo {volume} {19}},\ \bibinfo {pages} {1043}
  (\bibinfo {year} {2007})}\BibitemShut {NoStop}%
\bibitem [{\citenamefont {Wang}\ \emph {et~al.}(2007)\citenamefont {Wang},
  \citenamefont {Carter}, \citenamefont {Lagutchev}, \citenamefont {Koh},
  \citenamefont {Seong}, \citenamefont {Cahill},\ and\ \citenamefont
  {Dlott}}]{WangScience2007}%
  \BibitemOpen
  \bibfield  {author} {\bibinfo {author} {\bibfnamefont {Z.}~\bibnamefont
  {Wang}}, \bibinfo {author} {\bibfnamefont {J.~A.}~\bibnamefont {Carter}},
  \bibinfo {author} {\bibfnamefont {A.}~\bibnamefont {Lagutchev}}, \bibinfo
  {author} {\bibfnamefont {Y.~K.}\ \bibnamefont {Koh}}, \bibinfo {author}
  {\bibfnamefont {N.-H.}~\bibnamefont {Seong}}, \bibinfo {author} {\bibfnamefont
  {D.~G.}~\bibnamefont {Cahill}}, \ and\ \bibinfo {author} {\bibfnamefont {D.~D.}\
  \bibnamefont {Dlott}},\ }\href@noop {} {\bibfield  {journal} {\bibinfo
  {journal} {Science}\ }\textbf {\bibinfo {volume} {317}},\ \bibinfo {pages}
  {787} (\bibinfo {year} {2007})}\BibitemShut {NoStop}%
\bibitem [{\citenamefont {Mahan}\ and\ \citenamefont {Sofo}(1996)}]{Mahan1996}%
  \BibitemOpen
  \bibfield  {author} {\bibinfo {author} {\bibfnamefont {G.~D.}\ \bibnamefont
  {Mahan}}\ and\ \bibinfo {author} {\bibfnamefont {J.~O.}\ \bibnamefont
  {Sofo}},\ }\href@noop {} {\bibfield  {journal} {\bibinfo  {journal}
  {Proc.~Natl.~Acad.~Sci.~U.S.A.}\ }\textbf {\bibinfo {volume} {93}},\ \bibinfo
  {pages} {7436} (\bibinfo {year} {1996})}\BibitemShut {NoStop}%
\bibitem [{\citenamefont {Humphrey}\ \emph {et~al.}(2002)\citenamefont
  {Humphrey}, \citenamefont {Newbury}, \citenamefont {Taylor},\ and\
  \citenamefont {Linke}}]{HumphreyPRL2002}%
  \BibitemOpen
  \bibfield  {author} {\bibinfo {author} {\bibfnamefont {T.~E.}\ \bibnamefont
  {Humphrey}}, \bibinfo {author} {\bibfnamefont {R.}~\bibnamefont {Newbury}},
  \bibinfo {author} {\bibfnamefont {R.~P.}\ \bibnamefont {Taylor}}, \ and\
  \bibinfo {author} {\bibfnamefont {H.}~\bibnamefont {Linke}},\ }\href
  {\doibase 10.1103/PhysRevLett.89.116801} {\bibfield  {journal} {\bibinfo
  {journal} {Phys. Rev. Lett.}\ }\textbf {\bibinfo {volume} {89}},\ \bibinfo
  {pages} {116801} (\bibinfo {year} {2002})}\BibitemShut {NoStop}%
\bibitem [{\citenamefont {den Broeck}(2007)}]{VandenBroeckACP2007}%
  \BibitemOpen
  \bibfield  {author} {\bibinfo {author} {\bibfnamefont {C.~V.}\ \bibnamefont
  {den Broeck}},\ }\href@noop {} {\bibfield  {journal} {\bibinfo  {journal}
  {Adv.~Chem.~Phys.}\ }\textbf {\bibinfo {volume} {135}},\ \bibinfo {pages}
  {189} (\bibinfo {year} {2007})}\BibitemShut {NoStop}%
\bibitem [{\citenamefont {Esposito}\ \emph {et~al.}(2009)\citenamefont
  {Esposito}, \citenamefont {Lindenberg},\ and\ \citenamefont {den
  Broeck}}]{EspositoEPL2009}%
  \BibitemOpen
  \bibfield  {author} {\bibinfo {author} {\bibfnamefont {M.}~\bibnamefont
  {Esposito}}, \bibinfo {author} {\bibfnamefont {K.}~\bibnamefont
  {Lindenberg}}, \ and\ \bibinfo {author} {\bibfnamefont {C.~V.}\ \bibnamefont
  {den Broeck}},\ }\href@noop {} {\bibfield  {journal} {\bibinfo  {journal}
  {Europhys.~Lett.}\ }\textbf {\bibinfo {volume} {85}},\ \bibinfo {pages}
  {60010} (\bibinfo {year} {2009})}\BibitemShut {NoStop}%
\bibitem [{\citenamefont {Nakpathomkun}\ \emph {et~al.}(2010)\citenamefont
  {Nakpathomkun}, \citenamefont {Xu},\ and\ \citenamefont
  {Linke}}]{NakpathomkunPRB2010}%
  \BibitemOpen
  \bibfield  {author} {\bibinfo {author} {\bibfnamefont {N.}~\bibnamefont
  {Nakpathomkun}}, \bibinfo {author} {\bibfnamefont {H.~Q.}\ \bibnamefont
  {Xu}}, \ and\ \bibinfo {author} {\bibfnamefont {H.}~\bibnamefont {Linke}},\
  }\href {\doibase 10.1103/PhysRevB.82.235428} {\bibfield  {journal} {\bibinfo
  {journal} {Phys. Rev. B}\ }\textbf {\bibinfo {volume} {82}},\ \bibinfo
  {pages} {235428} (\bibinfo {year} {2010})}\BibitemShut {NoStop}%
\bibitem [{\citenamefont {Karlstr\"om}\ \emph {et~al.}(2011)\citenamefont
  {Karlstr\"om}, \citenamefont {Pedersen}, \citenamefont {Samuelsson},\ and\
  \citenamefont {Wacker}}]{KarlstromPRB2011}%
  \BibitemOpen
  \bibfield  {author} {\bibinfo {author} {\bibfnamefont {O.}~\bibnamefont
  {Karlstr\"om}}, \bibinfo {author} {\bibfnamefont {J.~N.}\ \bibnamefont
  {Pedersen}}, \bibinfo {author} {\bibfnamefont {P.}~\bibnamefont
  {Samuelsson}}, \ and\ \bibinfo {author} {\bibfnamefont {A.}~\bibnamefont
  {Wacker}},\ }\href {\doibase 10.1103/PhysRevB.83.205412} {\bibfield
  {journal} {\bibinfo  {journal} {Phys. Rev. B}\ }\textbf {\bibinfo {volume}
  {83}},\ \bibinfo {pages} {205412} (\bibinfo {year} {2011})}\BibitemShut
  {NoStop}%
\bibitem [{Note1()}]{Note1}%
  \BibitemOpen
  \bibinfo {note} {P. Trocha and J. Barna{\'s}, arXiv:1108.2422v1; G.
  G{\'o}mez-Silva, O.{\'A}valos-Ovando, M. L. Ladr{\'o}n de Guevara and P. A.
  Orellana, arXiv:1108.4460v1.}\BibitemShut {Stop}%
\bibitem [{\citenamefont {Liu}\ \emph {et~al.}(2011)\citenamefont {Liu},
  \citenamefont {Zhang}, \citenamefont {Yang},\ and\ \citenamefont
  {Yang}}]{LiuNT2011}%
  \BibitemOpen
  \bibfield  {author} {\bibinfo {author} {\bibfnamefont {Y.~S.}\ \bibnamefont
  {Liu}}, \bibinfo {author} {\bibfnamefont {D.~B.}\ \bibnamefont {Zhang}},
  \bibinfo {author} {\bibfnamefont {X.~F.}\ \bibnamefont {Yang}}, \ and\
  \bibinfo {author} {\bibfnamefont {X.~F.}\ \bibnamefont {Yang}},\ }\href@noop
  {} {\bibfield  {journal} {\bibinfo  {journal} {Nanotechnology}\ }\textbf
  {\bibinfo {volume} {22}},\ \bibinfo {pages} {225201} (\bibinfo {year}
  {2011})}\BibitemShut {NoStop}%
\bibitem [{\citenamefont {Wierzbicki}\ and\ \citenamefont
  {Swirkowicz}(2011)}]{WierzbickiPRB2011}%
  \BibitemOpen
  \bibfield  {author} {\bibinfo {author} {\bibfnamefont {M.}~\bibnamefont
  {Wierzbicki}}\ and\ \bibinfo {author} {\bibfnamefont {R.}~\bibnamefont
  {Swirkowicz}},\ }\href {\doibase 10.1103/PhysRevB.84.075410} {\bibfield
  {journal} {\bibinfo  {journal} {Phys. Rev. B}\ }\textbf {\bibinfo {volume}
  {84}},\ \bibinfo {pages} {075410} (\bibinfo {year} {2011})}\BibitemShut
  {NoStop}%
\bibitem [{Note2()}]{Note2}%
  \BibitemOpen
  \bibinfo {note} {This definition of $\Gamma $ differs by a factor $2$ from
  the one used in Ref.~ \protect \rev@citealpnum
  {NakpathomkunPRB2010}.}\BibitemShut {Stop}%
\bibitem [{\citenamefont {Tsai}\ and\ \citenamefont
  {Simpson}(2002)}]{Tsaicpl2002}%
  \BibitemOpen
  \bibfield  {author} {\bibinfo {author} {\bibfnamefont {H.-H.~G.}\
  \bibnamefont {Tsai}}\ and\ \bibinfo {author} {\bibfnamefont {M.~C.}\
  \bibnamefont {Simpson}},\ }\href {\doibase DOI:
  10.1016/S0009-2614(01)01457-9} {\bibfield  {journal} {\bibinfo  {journal}
  {Chemical Physics Letters}\ }\textbf {\bibinfo {volume} {353}},\ \bibinfo
  {pages} {111 } (\bibinfo {year} {2002})}\BibitemShut {NoStop}%
\bibitem [{\citenamefont {Humphrey}\ \emph {et~al.}(2005)\citenamefont
  {Humphrey}, \citenamefont {O'Dwyer},\ and\ \citenamefont
  {Linke}}]{Humphrey2005}%
  \BibitemOpen
  \bibfield  {author} {\bibinfo {author} {\bibfnamefont {T.~E.}\ \bibnamefont
  {Humphrey}}, \bibinfo {author} {\bibfnamefont {M.~F.}\ \bibnamefont
  {O'Dwyer}}, \ and\ \bibinfo {author} {\bibfnamefont {H.}~\bibnamefont
  {Linke}},\ }\href@noop {} {\bibfield  {journal} {\bibinfo  {journal}
  {J.~Phys.~D:~Appl.~Phys.}\ }\textbf {\bibinfo {volume} {38}},\ \bibinfo
  {pages} {2051} (\bibinfo {year} {2005})}\BibitemShut {NoStop}%
\bibitem [{\citenamefont {Cardamone}\ \emph {et~al.}(2006)\citenamefont
  {Cardamone}, \citenamefont {Stafford},\ and\ \citenamefont
  {Mazumdar}}]{CardamoneNanoLett2006}%
  \BibitemOpen
  \bibfield  {author} {\bibinfo {author} {\bibfnamefont {D.~M.}\ \bibnamefont
  {Cardamone}}, \bibinfo {author} {\bibfnamefont {C.~A.}\ \bibnamefont
  {Stafford}}, \ and\ \bibinfo {author} {\bibfnamefont {S.}~\bibnamefont
  {Mazumdar}},\ }\href@noop {} {\bibfield  {journal} {\bibinfo  {journal} {Nano
  Letters}\ }\textbf {\bibinfo {volume} {6}},\ \bibinfo {pages} {2422}
  (\bibinfo {year} {2006})}\BibitemShut {NoStop}%
\bibitem [{\citenamefont {Ke}\ \emph {et~al.}(2008)\citenamefont {Ke},
  \citenamefont {Yang},\ and\ \citenamefont {Baranger}}]{KeNanoLett2008}%
  \BibitemOpen
  \bibfield  {author} {\bibinfo {author} {\bibfnamefont {S.-H.}\ \bibnamefont
  {Ke}}, \bibinfo {author} {\bibfnamefont {W.}~\bibnamefont {Yang}}, \ and\
  \bibinfo {author} {\bibfnamefont {H.~U.}\ \bibnamefont {Baranger}},\
  }\href@noop {} {\bibfield  {journal} {\bibinfo  {journal} {Nano Letters}\
  }\textbf {\bibinfo {volume} {8}},\ \bibinfo {pages} {3257} (\bibinfo {year}
  {2008})}\BibitemShut {NoStop}%
\bibitem [{\citenamefont {Bergfield}\ and\ \citenamefont
  {Stafford}(2009)}]{BergfieldNanoLetters2009}%
  \BibitemOpen
  \bibfield  {author} {\bibinfo {author} {\bibfnamefont {J.~P.}\ \bibnamefont
  {Bergfield}}\ and\ \bibinfo {author} {\bibfnamefont {C.~A.}\ \bibnamefont
  {Stafford}},\ }\href@noop {} {\bibfield  {journal} {\bibinfo  {journal} {Nano
  Letters}\ }\textbf {\bibinfo {volume} {9}},\ \bibinfo {pages} {3072}
  (\bibinfo {year} {2009})}\BibitemShut {NoStop}%
\bibitem [{\citenamefont {Bergfield}\ \emph {et~al.}(2010)\citenamefont
  {Bergfield}, \citenamefont {Solis},\ and\ \citenamefont
  {Stafford}}]{BergfieldACSNano2010}%
  \BibitemOpen
  \bibfield  {author} {\bibinfo {author} {\bibfnamefont {J.~P.}\ \bibnamefont
  {Bergfield}}, \bibinfo {author} {\bibfnamefont {M.~A.}\ \bibnamefont
  {Solis}}, \ and\ \bibinfo {author} {\bibfnamefont {C.~A.}\ \bibnamefont
  {Stafford}},\ }\href@noop {} {\bibfield  {journal} {\bibinfo  {journal} {ACS
  Nano}\ }\textbf {\bibinfo {volume} {4}},\ \bibinfo {pages} {5314} (\bibinfo
  {year} {2010})}\BibitemShut {NoStop}%
\bibitem [{\citenamefont {{Nakanishi}}\ and\ \citenamefont
  {{Kato}}(2007)}]{Nakanishi_arXiv2007}%
  \BibitemOpen
  \bibfield  {author} {\bibinfo {author} {\bibfnamefont {T.}~\bibnamefont
  {{Nakanishi}}}\ and\ \bibinfo {author} {\bibfnamefont {T.}~\bibnamefont
  {{Kato}}},\ }\href {\doibase 10.1143/JPSJ.76.034715} {\bibfield  {journal}
  {\bibinfo  {journal} {Journal of the Physical Society of Japan}\ }\textbf
  {\bibinfo {volume} {76}},\ \bibinfo {pages} {034715} (\bibinfo {year}
  {2007})}\BibitemShut {NoStop}%
\bibitem [{\citenamefont {Finch}\ \emph {et~al.}(2009)\citenamefont {Finch},
  \citenamefont {Garc\'\i{}a-Su\'arez},\ and\ \citenamefont
  {Lambert}}]{FinchPRB2009}%
  \BibitemOpen
  \bibfield  {author} {\bibinfo {author} {\bibfnamefont {C.~M.}\ \bibnamefont
  {Finch}}, \bibinfo {author} {\bibfnamefont {V.~M.}\ \bibnamefont
  {Garc\'\i{}a-Su\'arez}}, \ and\ \bibinfo {author} {\bibfnamefont {C.~J.}\
  \bibnamefont {Lambert}},\ }\href {\doibase 10.1103/PhysRevB.79.033405}
  {\bibfield  {journal} {\bibinfo  {journal} {Phys. Rev. B}\ }\textbf {\bibinfo
  {volume} {79}},\ \bibinfo {pages} {033405} (\bibinfo {year}
  {2009})}\BibitemShut {NoStop}%
\bibitem [{\citenamefont {Markussen}\ \emph {et~al.}(2010)\citenamefont
  {Markussen}, \citenamefont {Stadler},\ and\ \citenamefont
  {Thygesen}}]{MarkussenNL2010}%
  \BibitemOpen
  \bibfield  {author} {\bibinfo {author} {\bibfnamefont {T.}~\bibnamefont
  {Markussen}}, \bibinfo {author} {\bibfnamefont {R.}~\bibnamefont {Stadler}},
  \ and\ \bibinfo {author} {\bibfnamefont {K.~S.}\ \bibnamefont {Thygesen}},\
  }\href@noop {} {\bibfield  {journal} {\bibinfo  {journal} {Nano Letters}\
  }\textbf {\bibinfo {volume} {10}},\ \bibinfo {pages} {4260} (\bibinfo {year}
  {2010})}\BibitemShut {NoStop}%
\bibitem [{\citenamefont {Markussen}\ \emph {et~al.}(2011)\citenamefont
  {Markussen}, \citenamefont {Stadler},\ and\ \citenamefont
  {Thygesen}}]{MarkussenPCCP2011}%
  \BibitemOpen
  \bibfield  {author} {\bibinfo {author} {\bibfnamefont {T.}~\bibnamefont
  {Markussen}}, \bibinfo {author} {\bibfnamefont {R.}~\bibnamefont {Stadler}},
  \ and\ \bibinfo {author} {\bibfnamefont {K.~S.}\ \bibnamefont {Thygesen}},\
  }\href@noop {} {\bibfield  {journal} {\bibinfo  {journal} {Phys. Chem. Cehm.
  Phys}\ }\textbf {\bibinfo {volume} {13}},\ \bibinfo {pages} {14311} (\bibinfo
  {year} {2011})}\BibitemShut {NoStop}%
\bibitem [{\citenamefont {Stadler}\ and\ \citenamefont
  {Markussen}(2011)}]{StadlerarXiv}%
  \BibitemOpen
  \bibfield  {author} {\bibinfo {author} {\bibfnamefont {R.}~\bibnamefont
  {Stadler}}\ and\ \bibinfo {author} {\bibfnamefont {T.}~\bibnamefont
  {Markussen}},\ }\href@noop {} {\bibfield  {journal} {\bibinfo  {journal}
  {arXiv:1106.3661v1}\ } (\bibinfo {year} {2011})}\BibitemShut {NoStop}%
\bibitem [{\citenamefont {Scheibner}\ \emph {et~al.}(2005)\citenamefont
  {Scheibner}, \citenamefont {Buhmann}, \citenamefont {Reuter}, \citenamefont
  {Kiselev},\ and\ \citenamefont {Molenkamp}}]{ScheibnerPRL2005}%
  \BibitemOpen
  \bibfield  {author} {\bibinfo {author} {\bibfnamefont {R.}~\bibnamefont
  {Scheibner}}, \bibinfo {author} {\bibfnamefont {H.}~\bibnamefont {Buhmann}},
  \bibinfo {author} {\bibfnamefont {D.}~\bibnamefont {Reuter}}, \bibinfo
  {author} {\bibfnamefont {M.~N.}\ \bibnamefont {Kiselev}}, \ and\ \bibinfo
  {author} {\bibfnamefont {L.~W.}\ \bibnamefont {Molenkamp}},\ }\href {\doibase
  10.1103/PhysRevLett.95.176602} {\bibfield  {journal} {\bibinfo  {journal}
  {Phys. Rev. Lett.}\ }\textbf {\bibinfo {volume} {95}},\ \bibinfo {pages}
  {176602} (\bibinfo {year} {2005})}\BibitemShut {NoStop}%
\bibitem [{\citenamefont {Hoffmann}\ \emph {et~al.}(2009)\citenamefont
  {Hoffmann}, \citenamefont {Nilsson}, \citenamefont {Matthews}, \citenamefont
  {Nakpathomkun}, \citenamefont {Persson}, \citenamefont {Samuelson},\ and\
  \citenamefont {Linke}}]{HoffmannNanoLett2009}%
  \BibitemOpen
  \bibfield  {author} {\bibinfo {author} {\bibfnamefont {E.~A.}\ \bibnamefont
  {Hoffmann}}, \bibinfo {author} {\bibfnamefont {H.~A.}\ \bibnamefont
  {Nilsson}}, \bibinfo {author} {\bibfnamefont {J.~E.}\ \bibnamefont
  {Matthews}}, \bibinfo {author} {\bibfnamefont {N.}~\bibnamefont
  {Nakpathomkun}}, \bibinfo {author} {\bibfnamefont {A.~I.}\ \bibnamefont
  {Persson}}, \bibinfo {author} {\bibfnamefont {L.}~\bibnamefont {Samuelson}},
  \ and\ \bibinfo {author} {\bibfnamefont {H.}~\bibnamefont {Linke}},\
  }\href@noop {} {\bibfield  {journal} {\bibinfo  {journal} {Nano Letters}\
  }\textbf {\bibinfo {volume} {9}},\ \bibinfo {pages} {779} (\bibinfo {year}
  {2009})}\BibitemShut {NoStop}%
\bibitem [{\citenamefont {Nilsson}\ \emph {et~al.}(2010)\citenamefont
  {Nilsson}, \citenamefont {Karlstr\"om}, \citenamefont {Larsson},
  \citenamefont {Caroff}, \citenamefont {Pedersen}, \citenamefont {Samuelson},
  \citenamefont {Wacker}, \citenamefont {Wernersson},\ and\ \citenamefont
  {Xu}}]{NilssonPRL2010}%
  \BibitemOpen
  \bibfield  {author} {\bibinfo {author} {\bibfnamefont {H.~A.}\ \bibnamefont
  {Nilsson}}, \bibinfo {author} {\bibfnamefont {O.}~\bibnamefont
  {Karlstr\"om}}, \bibinfo {author} {\bibfnamefont {M.}~\bibnamefont
  {Larsson}}, \bibinfo {author} {\bibfnamefont {P.}~\bibnamefont {Caroff}},
  \bibinfo {author} {\bibfnamefont {J.~N.}\ \bibnamefont {Pedersen}}, \bibinfo
  {author} {\bibfnamefont {L.}~\bibnamefont {Samuelson}}, \bibinfo {author}
  {\bibfnamefont {A.}~\bibnamefont {Wacker}}, \bibinfo {author} {\bibfnamefont
  {L.-E.}\ \bibnamefont {Wernersson}}, \ and\ \bibinfo {author} {\bibfnamefont
  {H.~Q.}\ \bibnamefont {Xu}},\ }\href {\doibase
  10.1103/PhysRevLett.104.186804} {\bibfield  {journal} {\bibinfo  {journal}
  {Phys. Rev. Lett.}\ }\textbf {\bibinfo {volume} {104}},\ \bibinfo {pages}
  {186804} (\bibinfo {year} {2010})}\BibitemShut {NoStop}%
\bibitem [{\citenamefont {Nilsson}\ \emph {et~al.}(2009)\citenamefont
  {Nilsson}, \citenamefont {Caroff}, \citenamefont {Thelander}, \citenamefont
  {Larsson}, \citenamefont {Wagner}, \citenamefont {Wernersson}, \citenamefont
  {Samuelson},\ and\ \citenamefont {Xu}}]{NilssonNL2009}%
  \BibitemOpen
  \bibfield  {author} {\bibinfo {author} {\bibfnamefont {H.~A.}\ \bibnamefont
  {Nilsson}}, \bibinfo {author} {\bibfnamefont {P.}~\bibnamefont {Caroff}},
  \bibinfo {author} {\bibfnamefont {C.}~\bibnamefont {Thelander}}, \bibinfo
  {author} {\bibfnamefont {M.}~\bibnamefont {Larsson}}, \bibinfo {author}
  {\bibfnamefont {J.~B.}\ \bibnamefont {Wagner}}, \bibinfo {author}
  {\bibfnamefont {L.}~\bibnamefont {Wernersson}}, \bibinfo {author}
  {\bibfnamefont {L.}~\bibnamefont {Samuelson}}, \ and\ \bibinfo {author}
  {\bibfnamefont {H.~Q.}\ \bibnamefont {Xu}},\ }\href@noop {} {\bibfield
  {journal} {\bibinfo  {journal} {Nano Letters}\ }\textbf {\bibinfo {volume}
  {9}},\ \bibinfo {pages} {3151} (\bibinfo {year} {2009})}\BibitemShut
  {NoStop}%
\bibitem [{\citenamefont {Walsh}\ \emph {et~al.}(2006)\citenamefont {Walsh},
  \citenamefont {Gordon}, \citenamefont {Officer},\ and\ \citenamefont
  {Campbell}}]{WalshTheochem2006}%
  \BibitemOpen
  \bibfield  {author} {\bibinfo {author} {\bibfnamefont {P.~J.}\ \bibnamefont
  {Walsh}}, \bibinfo {author} {\bibfnamefont {K.~C.}\ \bibnamefont {Gordon}},
  \bibinfo {author} {\bibfnamefont {D.~L.}\ \bibnamefont {Officer}}, \ and\
  \bibinfo {author} {\bibfnamefont {W.~M.}\ \bibnamefont {Campbell}},\
  }\href@noop {} {\bibfield  {journal} {\bibinfo  {journal} {J.~Mol.~Struct.
  (THEOCHEM)}\ }\textbf {\bibinfo {volume} {759}},\ \bibinfo {pages} {17}
  (\bibinfo {year} {2006})}\BibitemShut {NoStop}%
\bibitem [{\citenamefont {Xue}\ \emph {et~al.}(2001)\citenamefont {Xue},
  \citenamefont {Datta},\ and\ \citenamefont {Ratner}}]{XueJCP2001}%
  \BibitemOpen
  \bibfield  {author} {\bibinfo {author} {\bibfnamefont {Y.}~\bibnamefont
  {Xue}}, \bibinfo {author} {\bibfnamefont {S.}~\bibnamefont {Datta}}, \ and\
  \bibinfo {author} {\bibfnamefont {M.~A.}\ \bibnamefont {Ratner}},\
  }\href@noop {} {\bibfield  {journal} {\bibinfo  {journal} {Journal of
  Chemical Physics}\ }\textbf {\bibinfo {volume} {115}},\ \bibinfo {pages}
  {4292} (\bibinfo {year} {2001})}\BibitemShut {NoStop}%
\bibitem [{\citenamefont {Xue}\ \emph {et~al.}(2002)\citenamefont {Xue},
  \citenamefont {Datta},\ and\ \citenamefont {Ratner}}]{XueCP2002}%
  \BibitemOpen
  \bibfield  {author} {\bibinfo {author} {\bibfnamefont {Y.}~\bibnamefont
  {Xue}}, \bibinfo {author} {\bibfnamefont {S.}~\bibnamefont {Datta}}, \ and\
  \bibinfo {author} {\bibfnamefont {M.~A.}\ \bibnamefont {Ratner}},\
  }\href@noop {} {\bibfield  {journal} {\bibinfo  {journal} {Chemical Physics}\
  }\textbf {\bibinfo {volume} {281}},\ \bibinfo {pages} {151} (\bibinfo {year}
  {2002})}\BibitemShut {NoStop}%
\bibitem [{\citenamefont {Tian}\ \emph {et~al.}(1998)\citenamefont {Tian},
  \citenamefont {Datta}, \citenamefont {Hong}, \citenamefont {Reifenberger},
  \citenamefont {Henderson},\ and\ \citenamefont {Kubiak}}]{TianJCP1998}%
  \BibitemOpen
  \bibfield  {author} {\bibinfo {author} {\bibfnamefont {W.}~\bibnamefont
  {Tian}}, \bibinfo {author} {\bibfnamefont {S.}~\bibnamefont {Datta}},
  \bibinfo {author} {\bibfnamefont {S.}~\bibnamefont {Hong}}, \bibinfo {author}
  {\bibfnamefont {R.}~\bibnamefont {Reifenberger}}, \bibinfo {author}
  {\bibfnamefont {J.~I.}\ \bibnamefont {Henderson}}, \ and\ \bibinfo {author}
  {\bibfnamefont {C.~P.}\ \bibnamefont {Kubiak}},\ }\href@noop {} {\bibfield
  {journal} {\bibinfo  {journal} {J.~Chem.~Phys.}\ }\textbf {\bibinfo {volume}
  {109}},\ \bibinfo {pages} {2874} (\bibinfo {year} {1998})}\BibitemShut
  {NoStop}%
\bibitem [{\citenamefont {Mingo}(2006)}]{MingoPRB2006}%
  \BibitemOpen
  \bibfield  {author} {\bibinfo {author} {\bibfnamefont {N.}~\bibnamefont
  {Mingo}},\ }\href {\doibase 10.1103/PhysRevB.74.125402} {\bibfield  {journal}
  {\bibinfo  {journal} {Phys. Rev. B}\ }\textbf {\bibinfo {volume} {74}},\
  \bibinfo {pages} {125402} (\bibinfo {year} {2006})}\BibitemShut {NoStop}%
\bibitem [{\citenamefont {Jarzecki}\ \emph {et~al.}(1997)\citenamefont
  {Jarzecki}, \citenamefont {Kozlowski}, \citenamefont {Pulay}, \citenamefont
  {Ye},\ and\ \citenamefont {Li}}]{Jarzecki1997}%
  \BibitemOpen
  \bibfield  {author} {\bibinfo {author} {\bibfnamefont {A.~A.}\ \bibnamefont
  {Jarzecki}}, \bibinfo {author} {\bibfnamefont {P.~M.}\ \bibnamefont
  {Kozlowski}}, \bibinfo {author} {\bibfnamefont {P.}~\bibnamefont {Pulay}},
  \bibinfo {author} {\bibfnamefont {B.-H.}\ \bibnamefont {Ye}}, \ and\ \bibinfo
  {author} {\bibfnamefont {X.-Y.}\ \bibnamefont {Li}},\ }\href@noop {}
  {\bibfield  {journal} {\bibinfo  {journal} {Spectrochimica~Acta~Part~A}\
  }\textbf {\bibinfo {volume} {53}},\ \bibinfo {pages} {1195} (\bibinfo {year}
  {1997})}\BibitemShut {NoStop}%
\bibitem [{\citenamefont {Pedersen}\ and\ \citenamefont
  {Wacker}(2005)}]{PedersenPRB2005a}%
  \BibitemOpen
  \bibfield  {author} {\bibinfo {author} {\bibfnamefont {J.~N.}\ \bibnamefont
  {Pedersen}}\ and\ \bibinfo {author} {\bibfnamefont {A.}~\bibnamefont
  {Wacker}},\ }\href@noop {} {\bibfield  {journal} {\bibinfo  {journal}
  {Phys.~Rev.~B}\ }\textbf {\bibinfo {volume} {72}},\ \bibinfo {pages} {195330}
  (\bibinfo {year} {2005})}\BibitemShut {NoStop}%
\end{thebibliography}

%

\end{document}